\documentclass{PoS}
\usepackage{amsmath,amssymb,graphicx}

\title{Worm algorithm for the $O(2N)$ Gross-Neveu model}

\ShortTitle{Worm algorithm for the GN model}

\author{\speaker{Vidushi Maillart}\\
        Albert Einstein Center for Fundamental Physics, Institute for
        Theoretical Physics, \\
        Sidlerstrasse 5, CH-3012 Bern, Switzerland \\
        Institute for Theoretical Physics, University of Regensburg, 93040 Regensburg, Germany\\
        E-mail: \email{vidushi.maillart@itp.unibe.ch}}

\author{Urs Wenger\\
        Albert Einstein Center for Fundamental Physics, Institute for
        Theoretical Physics,\\
        Sidlerstrasse 5, CH-3012 Bern, Switzerland\\
        E-mail: \email{wenger@itp.unibe.ch}}

      \abstract{We study the lattice $O(2N)$ Gross-Neveu model with
        Wilson fermions in the fermion loop formulation. Employing
        a worm algorithm for an open fermionic string, we simulate fluctuating
        topological boundary conditions and use them to tune the system
        to the critical point. We show how the worm algorithm can be
        extended to sample correlation functions of bound states
        involving an arbitrary number of Majorana fermions and present
        first results.}

\FullConference{The XXVIII International Symposium on Lattice Field
  Theory, Lattice2010\\ June 14-19, 2010\\ Villasimius, Italy}

\begin{document}

\section{Introduction}
It is well known that fermionic degrees of freedom are difficult to
simulate on the lattice due to their Grassmannian nature. The fermion
matrix determinant, obtained after integrating out the fermion fields,
remains the major bottleneck of Monte-Carlo simulations, since it is
highly non-local and connects all the degrees of freedom. Each
Monte-Carlo update step changes the fermion matrix and requires a new
calculation of the determinant. On the other hand, if an algorithm can
update the determinant by a local procedure, then a significant gain
in efficiency might be achievable. However, simulations of fermions
usually suffer from critical slowing down when the fermion correlation
length diverges, i.e.~when the fermions become massless, and it is by
no means obvious whether a local algorithm can be constructed to
circumvent this problem.

An interesting attempt to deal with the problems related to simulating
fermions on the lattice has recently been suggested in
\cite{Wenger:2008tq,Wenger:2009mi} in the context of the $O(2N)$
Gross-Neveu (GN) model. It is based on reformulating the model in
terms of closed fermion loops
\cite{Stamatescu:1980br,Gattringer:1998cd,Wolff:2007ip}.  Introducing
an open fermionic string, an algorithm can be devised for which
critical slowing is essentially absent. It follows the spirit of
Prokof'ev and Svistunov's worm algorithm \cite{Prokof'ev:2001zz} and
makes use of the fact that a global update of the closed fermion loops
can be obtained by locally updating the open fermionic string.  The
open string corresponds to the insertion of a Majorana fermion pair
and directly samples the two-point correlation function. In this way
the configurations are updated on all length scales up to the
correlation length, and this eventually guarantees the absence of
critical slowing down.  Moreover, the algorithm also allows
simulations directly in the massless limit
\cite{Wenger:2008tq,Wenger:2009mi} and provides direct access to the
critical point via ratios of partition functions \cite{Bar:2009yq}.

The algorithm has been successfully applied to simulate free fermions
in two dimensions, but it has proven
equally successfull also in its application to strongly interacting
fermions, e.g.~in the Schwinger model in the strong coupling limit
\cite{Wenger:2008tq,Wenger:2009mi} and in supersymmetric quantum
mechanics \cite{Baumgartner:2011cm}. Here we report on the application
of the worm algorithm to another two-dimensional system of interacting
fermions - the $O(2N)$ Gross-Neveu model. We demonstrate how a simple
modification of the worm algorithm can be used to measure the fermion
bound state spectrum and we present preliminary results for $N=1$, in
which case the GN model corresponds to the Thirring model.

\section{Fermion loop formulation of the Gross Neveu model}
We consider the two-dimensional $O(2N)$-symmetric GN model
\cite{Gross:1974jv} described by the Lagrangian
\begin{equation} \label{eq:GN}
\mathcal{L} = \sum_{i=1}^N \overline{\psi}_i (\gamma_\mu \partial_\mu
+ m) \psi_i - \frac{g^2}{2} (\sum_{i=1}^N \overline{\psi}_i \psi_i)^2 \,.
\end{equation}
This is a relativistic quantum field theory of $N$ self-interacting
Dirac fermion fields. For $N=1$ the $O(2)$ GN model is equivalent to
the massive Thirring model as $(\overline{\psi} \psi)^2 =
{\frac{1}{4}}(\overline{\psi}\gamma_\mu\psi)(\overline{\psi}\gamma_\mu\psi)$. It
has a pseudoscalar bosonic fermion-antifermion bound state
\cite{Dashen:1975hd} and is especially interesting due to its
equivalence to the sine-Gordon model \cite{Coleman:1974bu}, in which
the boson is the fundamental particle and the fermion emerges as a
soliton solution.

To make the model amenable for the fermion loop algorithm in
\cite{Wenger:2008tq}, we decompose the complex Dirac fermion fields
$\psi_j$ and $\overline{\psi}_j$ into real Majorana fields $\xi_{2j}$
and $\xi_{2j+1}$ according to $\psi_j = {(\xi_{2j} + i \xi_{2j+1})/\sqrt{2}}$ and $\overline{\psi}_j = (\overline{\xi}_{2j} -
i \overline{\xi}_{2j+1}) \sqrt{2}$.  The Lagrangian density can then
be written as
\begin{equation} \label{eq:L}
\mathcal{L} = \frac{1}{2} \sum_{i=1}^{2N} \overline{\xi}_i (\gamma_\mu \partial_\mu + m) \xi_i - \frac{g^2}{8} \left(\sum_{i=1}^2 \overline{\xi}_i \xi_i\right)^2 \,.
\end{equation}
Note that the $O(2N)$ symmetry is now obvious, as \eqref{eq:L} is
invariant under rotations of the Majorana fields in flavour space.
Discretising the action with Wilson fermions gives
\begin{equation} \label{eq:S}
\mathrm{S} =  \frac{1}{2}\sum_x\sum_{i=1}^{2N} \varphi
\overline{\xi}_i(x) \xi_i(x) -  \frac{g^2}{8}
\sum_x\left(\sum_{i=1}^{2N} \overline{\xi}_i(x) \xi_i(x)\right)^2 -
\sum_{i=1}^{2N} \sum_{x,\mu} (\overline{\xi}_i (x) P(\mu) \xi_i (x+ \hat{\mu})) \,, 
\end{equation}
where $\varphi = (2 + m)$ and $P(\pm \mu) = \frac{1}{2} (1 \mp
\gamma_\mu)$. When expanding the Boltzmann factor in the partition
function to all orders, terms quadratic or higher order in each field
component vanish due to the nilpotency of the Grassmann
fields. Restricting now to $N=1$ for simplicity and keeping only
non-trivial terms, the partition function can be written as
\begin{equation} \label{eq:Z}
\begin{split}
\mathrm{Z} = \int \mathcal{D} \xi &  \prod_x \left(1 - \frac{\varphi}{2} \left(\sum_{i=1}^2 \overline{\xi}_i (x) \xi_i (x)\right) + \frac{(\varphi^2 + g^2)}{4} \overline{\xi}_1(x) \xi_1(x) \overline{\xi}_2 (x)\xi_2(x) \right) \\ 
 & \times  \prod_{x,\mu} \, \prod_{i=1}^2\left(1 + \overline{\xi}_i (x) P(\hat{\mu}) \xi_i
   (x+ \hat{\mu}) \right) 
\end{split}
\end{equation}
The integration measure is saturated site by site by combinations
$\overline{\xi}_i \xi_i$, and it is straightforward to identify the
non-vanishing contributions to the partition function.  They are
characterised by the fact that for a particular Majorana flavour
either two adjacent hopping terms and no monomer terms, or one of the
monomer terms, but no hopping terms are present at a given site. This
results in the constraint that for each flavour only closed,
non-intersecting fermion loops survive the Grassmann
integration. Furthermore, the loops are non-backtracking due to the
orthogonality of the projectors, viz.~$P(+\mu)P(-\mu) = 0$.  Thus, the
partition function is a sum over all possible combinations of two
different species of loops (corresponding to the two Majorana
flavours). The generalisation to an arbitrary number of $N$ Dirac
fields is straightforward -- the number of species of loops involved
is simply equal to $2N$, the number of Majorana fermions.

\section{Worm algorithm for Majorana fermions}
\label{sec:single_fermion_algorithm} 
In order to generate configurations of closed loops we employ a
variant of the algorithm of Prokof'ev and Svistunov
\cite{Prokof'ev:2001zz}. Here we explain the main ideas of the open
fermionic string (``worm'') algorithm in a few schematic steps and
point out the modifications we have introduced to increase
efficiency. Further details can be found in
\cite{Prokof'ev:2001zz,Wenger:2008tq}.

A peculiar feature of the worm algorithm is that the fermion
correlation function is measured during the update procedure. This is
due to the fact that the insertion of the open fermionic string, which
is used to update the loop configuration, corresponds to the insertion
of a pair of Majorana fermions $\xi_i(x) \overline{\xi}_i(y)$ of
flavour $i$ at positions $x$ and $y$, respectively. In the path
integral formalism this is equivalent to the correlation function
 \begin{equation}
	G_i(x,y) = \langle \xi_i(x) \overline{\xi}_i(y)\rangle =
        \frac{1}{Z}\int \mathcal{D} \xi  \,
        \xi_i (x) \overline{\xi}_i (y) \, e^{- S} \,. 
 \end{equation}
The algorithm now proceeds by locally updating the ends of the open
string using a simple Metropolis procedure according to the weights
of the corresponding two-point function. The main steps are as follows:

\begin{itemize}
\item {\it Relocation step:} Given a closed loop configuration, choose
  a fermion flavour $i$ and a lattice site $x$ at random and place the
  head $\xi_i$ and the tail $\overline{\xi_i}$ of the worm on this
  site with a probability given by the weight ratio of the loop
  configurations before and after the step. If accepted, it gives a
  contribution to $G_i(x,x)$ unless a loop of species $i$ passes
  through $x$ in the original closed loop configuration, in which case
  the new configuration has no physical interpretation and gives no
  contribution to $G_i(x,x)$.
\item {\it Move step:} Choose a direction $\mu$ at random, and move
  the head of the worm to site $y = x+\hat \mu$. Add or delete a
  fermion bond between $x$ and $x+\hat \mu$ depending on whether the
  bond is empty or occupied. The resulting configuration gives a
  contribution to $G_i(x,x + \mu)$.
\item {\it Break-up/reconnect step:} In case the new site $x+\hat \mu$
  is already occupied by a fermion loop of flavour $i$, we still allow
  the move, although the corresponding configuration (cf.~middle plot
  in figure \ref{fig:breakup_reconnect}) is forbidden by the Pauli
  exclusion principle. Consequently, it does not contribute to the
  correlation function, but induces transitions between allowed
  configurations as indicated in the figure.
\begin{figure}[t!]	%[h!]
  \begin{center} 
\begin{minipage}[c]{3cm}
\centering
\includegraphics[width=0.6\textwidth]{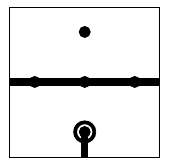}
\end{minipage}
\begin{minipage}[c]{0.7cm}
$\Longleftrightarrow$
\end{minipage}
\begin{minipage}[c]{3cm}
\centering
\includegraphics[width=0.6\textwidth]{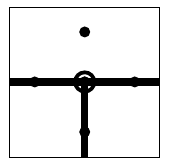}
\end{minipage}
\begin{minipage}[c]{0.7cm}
$\Longleftrightarrow$
\end{minipage}
\begin{minipage}[c]{3cm}
\centering
\includegraphics[width=0.6\textwidth]{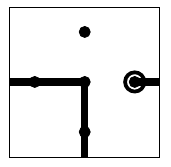}
\end{minipage}
  \end{center}
\vspace{-0.5cm}
\caption{Update moves for the break-up/reconnect step. Note that the
  configuration in the middle has no physical interpretation and hence
  gives no contribution to the correlation function.}
\label{fig:breakup_reconnect}
\end{figure}
\item {\it Removal step:} Once the head of the worm reaches its
  starting position, i.e.~head and tail meet again at site $x$, we may
  propose to remove head and tail. If accepted, we have a closed loop
  configuration contributing to the partition function.
\end{itemize}

We emphasise that the break-up/reconnect step is crucial for the
algorithm to work efficiently, since it allows the loops to be opened
and restructured. Especially close to the critical point, where loops
proliferate, this step gives the worm much more freedom to update the
configurations.

\section{Topological and fermionic boundary conditions} 
On a finite lattice with a periodic torus geometry, fermion loops can
wind around the lattice and the loop configurations can hence be
categorised into different homotopy classes depending on the number of
loop windings. For each Majorana fermion $\xi_i$ a two dimensional
vector $\vec{l}_i = (l_x, l_t)_i$ is assigned to each configuration to
account for the windings in space and time direction, respectively.
The components of $\vec{l}_i$ are either $0$ or $1$ corresponding to
an overall even or odd number of loop windings, respectively, in the
corresponding direction.  Configurations with different $\vec{l}_i$
contribute to separate partition functions $Z_{\vec{l}_1,
  \vec{l}_2,\ldots}$ with fixed topological boundary conditions
(b.c.). Since the open fermion string of the worm algorithm tunnels between
the configurations in the various homotopy classes, it samples all the
partition functions $Z_{\vec{l}_1, \vec{l}_2,\ldots}$. More importantly, 
the configurations in all homotopy classes are sampled with positive
weights relative to each other, i.e. the partition function ~$Z \equiv \sum_{\vec{l}_1,
  \vec{l}_2,\ldots} Z_{\vec{l}_1, \vec{l}_2, \ldots}$ 
corresponds to one with fluctuating topological b.c.,
but unspecified fermionic b.c. As a consequence, 
the b.c.~for the fermions can be chosen at the end of the 
simulation and all possible fermionic b.c.~can be studied {\it a 
posteriori}.

In order to encode the fermionic b.c.~we introduce a two dimensional
vector $\vec{\epsilon}_i$ in analogy to $\vec{l}_i$.  The components
of $\vec{\epsilon}_i$ are $0$ or $1$, and correspond to periodic or
anti-periodic b.c., respectively.  The partition function
$Z^{\vec{\epsilon}_1, \vec{\epsilon}_2,\ldots}$ for fixed
${\vec{\epsilon}_1, \vec{\epsilon}_2,\ldots}$ can now be written as a
linear combination of the partition functions $Z_{\vec{l}_1,
  \vec{l}_2,\ldots}$, e.g.~for $N=1$,
\begin{equation} \label{eq:partion_two_flav}
Z^{\vec{\epsilon_1}, \vec{\epsilon_2}} = 4 \, Z_{\vec{0}, \vec{0}}
 - 2 \sum_{\vec{l}_1} (-1)^{\vec{\epsilon_1} \vec{l_1}} \, Z_{\vec{l}_1,\vec{0}}
- 2 \sum_{\vec{l_2}} (-1)^{\vec{\epsilon_2} \vec{l_2}} \, Z_{\vec{0},\vec{l}_2}
+ \sum_{\vec{l}_1,\vec{l}_2}  (-1)^{(\vec{\epsilon_1} \vec{l_1} + \vec{\epsilon_2} \vec{l_2})} \,  Z_{\vec{l}_1, \vec{l}_2} \,.
\end{equation}
As shown in \cite{Bar:2009yq}, choosing periodic b.c.~for all fermions
in all directions, except antiperiodic in one direction for one single
fermion, e.g.~$\vec{\epsilon}_1=(0,1)$ and
$\vec{\epsilon}_{i>1}=(0,0)$, the corresponding partition function
$Z^{0100\ldots}$ vanishes at the massless, critical point, i.e.~when
the bare mass $m$ is equal to the critical mass $m_c$.  Hence, the
criterion can be used to determine $m_c$ for various couplings by
tuning the bare mass $m$ to the point where $Z^{0100\ldots}=0$.

\begin{figure}[t] %[h!]
\begin{center}
\includegraphics[width=7cm]{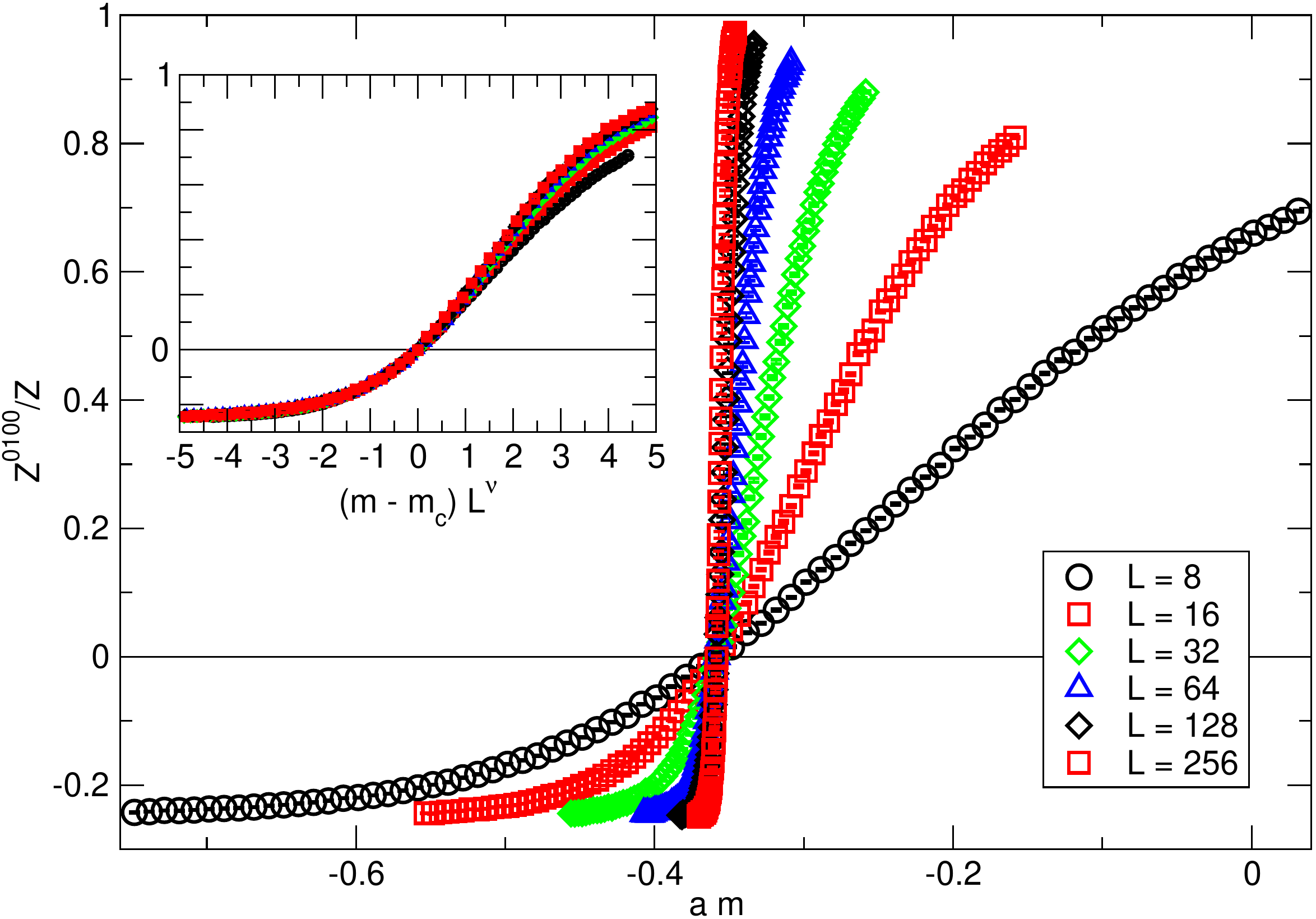}
\hfill
\includegraphics[width=7cm]{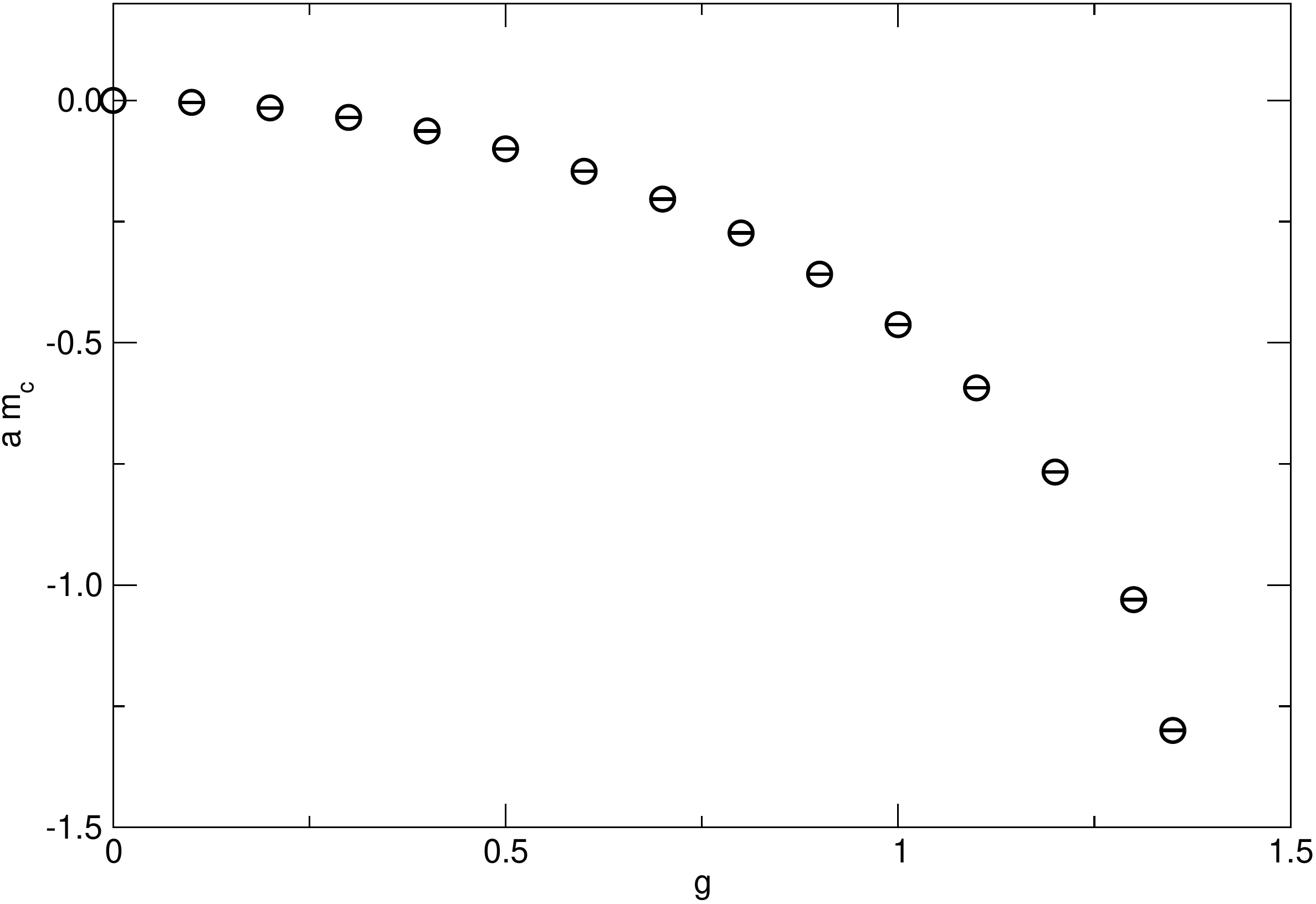}
\caption{The figure on the left shows the ratio $Z^{0100}/Z$ for the
  Thirring model at $g=0.9$ as a function of the bare mass $am$ on
  square lattices of size $L=8, 16, 32, 64, 128$ and $256$. The inset
  shows the finite size scaling. The right plot summarises our results
  for the critical mass $am_c$ as a function of the coupling $g$.}
\label{fig:m_crit}
\end{center}
\end{figure}

In figure \ref{fig:m_crit} we show the results of such a determination
for the Thirring model.
The plot on the left shows the partition function ratio $Z^{0100}/Z$
as a function of the bare mass $a m$ for square lattices
with $L=8, \ldots, 256$ at the coupling $g=0.9$. As the lattice size
increases the jump from $Z^{0100}/Z \sim 1$ to $Z^{0100}/Z \sim -0.25$
is more and more pronounced, so that the critical point $Z^{0100}= 0$
can be determined very precisely. Since the critical point corresponds
to a second order phase transition where the correlation length
diverges, one should find a corresponding universal finite size
scaling (FSS) behaviour. The inset in the left plot of figure
\ref{fig:m_crit} illustrates that this is indeed the case. There we
show the partition function ratios as a function of $(m - m_{c})L^\nu$,
and from the FSS we can determine the critical exponent $\nu$ and the
critical mass $a m_c$ in the thermodynamic limit $L\rightarrow \infty$
to a very high precision with a rather modest computational effort.
The right plot in figure \ref{fig:m_crit} summarises our results for the critical mass in
the thermodynamic limit as a function of the coupling $g$.

\section{Bound states of fermions}
It can easily be worked out that the correlation function of the
pseudoscalar bound state $\overline{\psi} \gamma_5 \psi$ in the
Thirring model is given by
\begin{equation} \label{eq:pseudo_bs}
  \langle\mathcal{O} (x) \mathcal{O} (y) \rangle = \langle \overline{\xi}_1 (x) \, \gamma_5 \, \xi_2 (x) \overline{\xi}_1 (y) \, \gamma_5 \, \xi_2 (y)\rangle \,.
 \end{equation}
In the loop formulation, this corresponds to two open fermion
strings, one for each Majorana fermion flavour, with common endpoints
at positions $x$ and $y$. In complete analogy to the single fermion
update using the fermionic 2-point correlation function, we can
update the configurations using the bosonic bound state correlation
function in eq.(\ref{eq:pseudo_bs}).  In practice, we insert two
instances of the bound state wave function $\mathcal{O} (x) =
\overline{\xi}_1 (x) \, \gamma_5 \, \xi_2 (x)$ into the system and
let them move around by employing again a local Metropolis
update. This procedure samples the bound state correlation function,
and at the same time updates the loop configuration of both fermion
flavours. In order for this to work efficiently, the break-up and
reconnection step described in section
\ref{sec:single_fermion_algorithm} is the crucial ingredient, since
otherwise the algorithm would be restricted to move the bound state
wave function only to sites where no fermion loop is
present. Obviously, this would become increasingly difficult towards
the critical point, where the fermion loops proliferate.

\begin{figure}[t!]
  \begin{center} 
	\includegraphics[width=7cm]{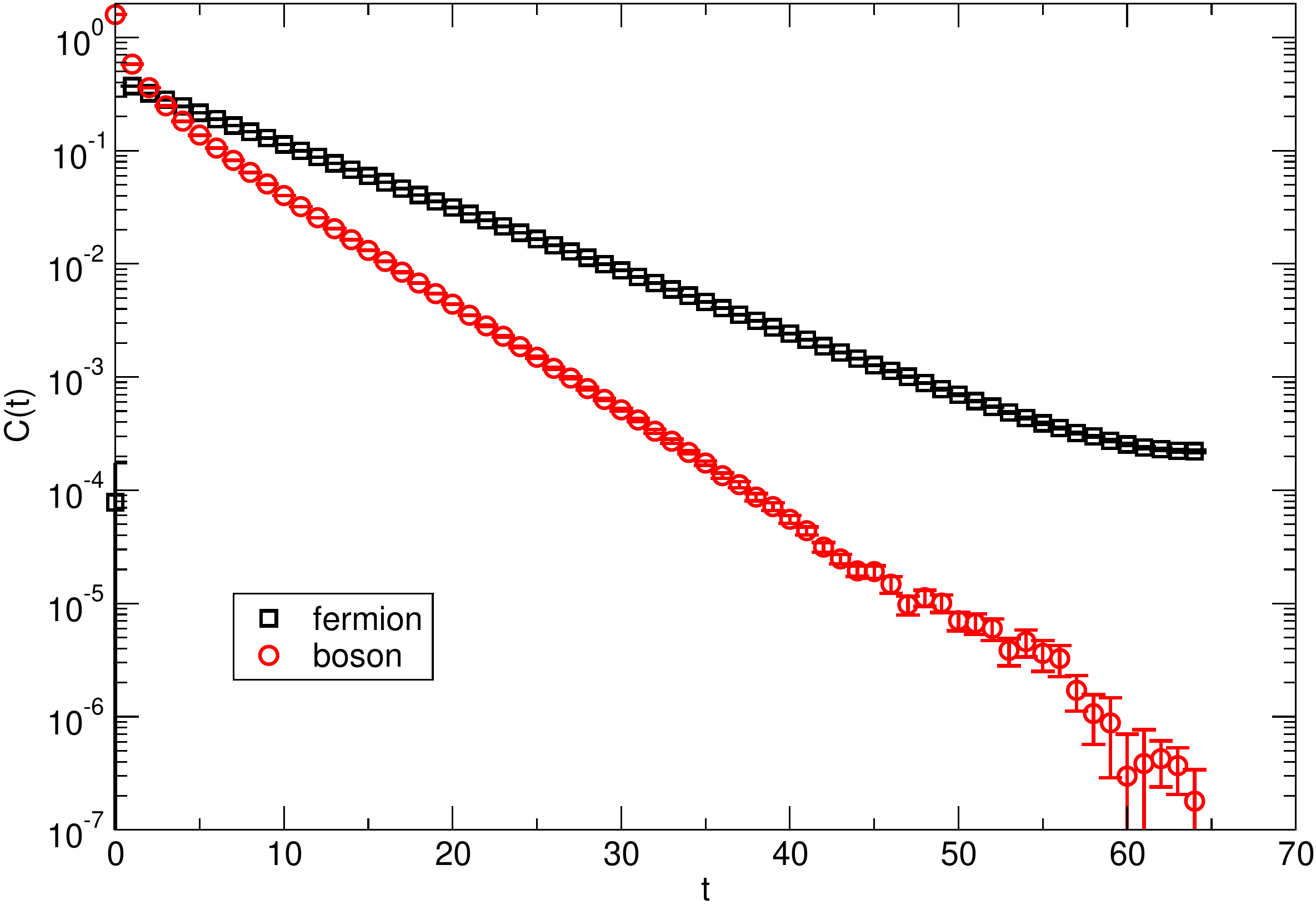}
        \hfill
	\includegraphics[width=7cm]{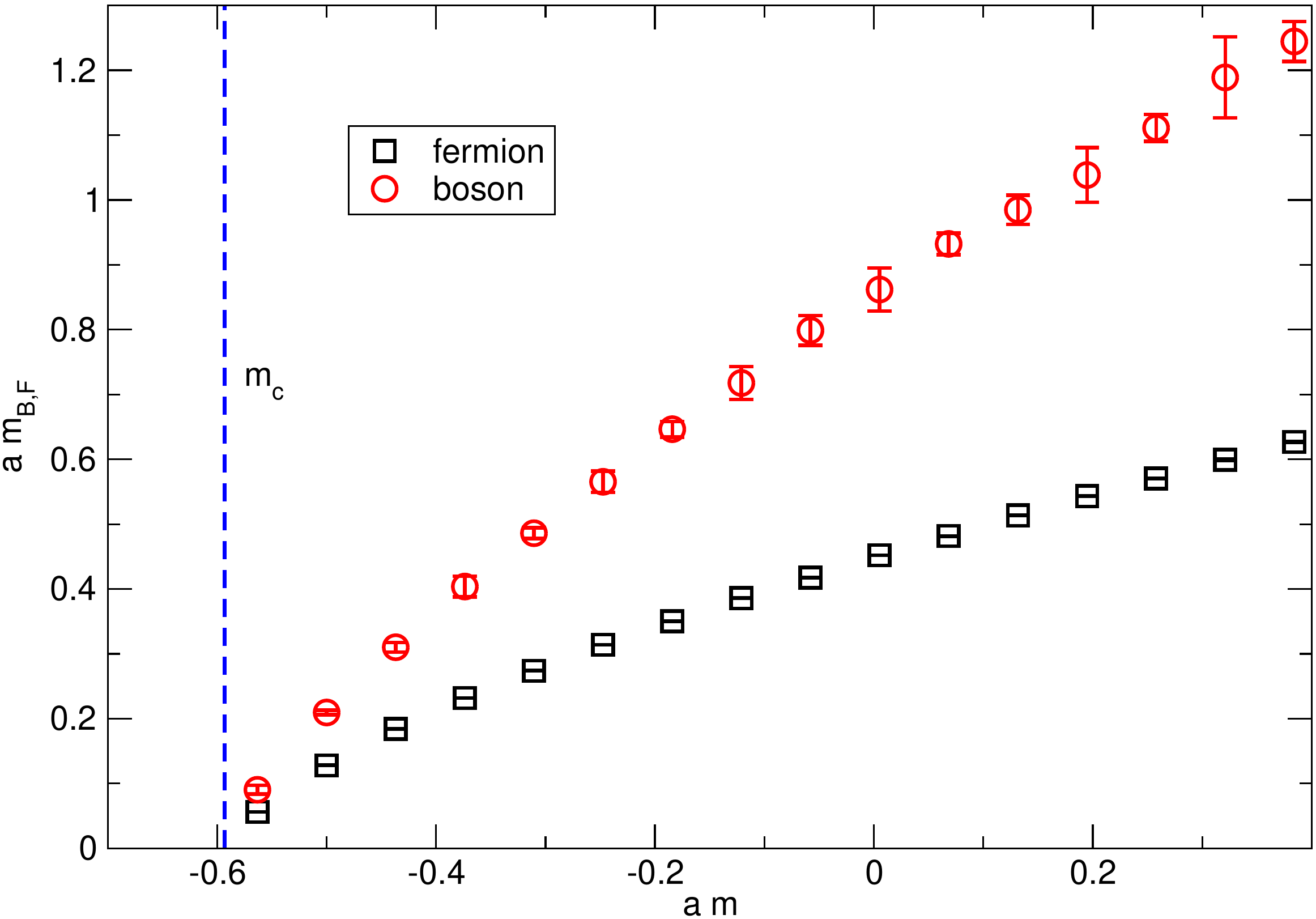}
	\caption{In the left plot the single fermion and bound state
          correlation functions are plotted. The right plot shows the
          fermion and boson masses as a function of the bare mass $am$
          for the Thirring model at $g=1.1$.}
	\label{fig:correlation_bs}	
  \end{center}
\end{figure}

The efficiency of the algorithm is illustrated in figure
\ref{fig:correlation_bs}.  In the left plot we show the single fermion
and the bosonic bound state correlation functions
at zero momentum obtained from a simulation of the Thirring model on a
$L=128$ lattice at coupling $g=1.1$. It is remarkable that in both
cases the signal can be followed over several orders of
magnitude. Consequently, the corresponding masses can be reliably
determined towards the continuum limit. This is illustrated in the
right plot of figure \ref{fig:correlation_bs} where we show the
fermion and boson masses versus the bare mass $am$.

\section{Conclusions and outlook}
Using Wilson's fermion discretisation, the path integral for the
$O(2N)$ Gross-Neveu (GN) model can be described on the lattice in
terms of interacting fermion loops. We discussed how the loop system
can efficiently be simulated using open fermion strings.  Single
fermion and bound state correlation functions are measured while
updating the system. In addition, the algorithm allows the direct
calculation of ratios of partition functions with arbitrary fermion
boundary conditions.  We have successfully implemented the fermion
loop algorithm for $N=1$, in which case the GN model is equivalent to
the Thirring model, and presented first preliminary results for the
determination of the critical point from the partition function
ratios. Moreover, we also presented first promising results for the
single fermion and the bound state masses.
Currently we are working on measuring these quantities for the massive
Thirring model in the continuum limit at various values of the
couplings, in order to compare the results to predictions based on the
equivalence of the model to the Sine-Gordon model.  The extension of
the algorithm to a larger number of fermions is interesting and rather
straightforward.

An obvious question to ask is whether and how the idea of updating an
open fermionic string can be put to use in the context of gauged
fermions or in higher dimensions. Successfull attempts were so far
reported only in the strong coupling limit
\cite{Wenger:2008tq,Wenger:2009mi,Chandrasekharan:2010ik}, but there
are many other interesting and promising extensions
\cite{Chandrasekharan:2009wc} using worm-type algorithms, even in
connection with pure gauge theories \cite{Korzec:2010sh}.

\subsection*{Acknowledgements}
\noindent
This work has partly been supported by the European Union under grant 238353
(ITN STRONGnet).

%%%%%%%%%%%%%%%%%%%%%%%%%%%%%%%%%%%%%%%%%%%%%%%%%%
%\bibliographystyle{h-physrev4}
\bibliographystyle{JHEP} %if you use h-elsevier.bst
\bibliography{waftO2NGNm}

\end{document}